\documentclass[10pt,preprint]{aastex}

\shorttitle{Light Echo for SN 2006X} \shortauthors{Wang et al.}

\def\gsim{\;\lower4pt\hbox{${\buildrel\displaystyle >\over\sim}$}\;}
\def\lsim{\;\lower4pt\hbox{${\buildrel\displaystyle <\over\sim}$}\;}
\def\grls{\;\lower4pt\hbox{${\buildrel\displaystyle >\over <}$}\;}

\begin{document}

\title{The Detection of a Light Echo from the Type Ia \\ Supernova 2006X in M100}

\author{Xiaofeng Wang\altaffilmark{1,2}, Weidong Li\altaffilmark{1},
Alexei V. Filippenko\altaffilmark{1}, Ryan J. Foley\altaffilmark{1},
\\ Nathan Smith\altaffilmark{1}, and Lifan Wang\altaffilmark{3}}

\altaffiltext{1}{Department of Astronomy, University of California,
Berkeley, CA 94720-3411, USA; wangxf@astro.berkeley.edu .}
\altaffiltext{2}{Physics Department and Tsinghua Center for
Astrophysics (THCA), Tsinghua University, Beijing, 100084, China;
wang\_xf@mail.tsinghua.edu.cn .} \altaffiltext{3}{Physics
Department, Texas A\&M University, College Station, TX 77843 .}

\begin{abstract}

We report the discovery of a light echo (LE) from the Type Ia
supernova (SN) 2006X in the nearby galaxy M100. The presence of the
LE is supported by analysis of both the {\it Hubble Space Telescope
(HST)} Advanced Camera for Surveys (ACS) images and the Keck optical
spectrum that we obtained at $\sim$300~d after maximum brightness.
In the image procedure, both the radial-profile analysis and the
point-spread function (PSF) subtraction method resolve significant
excess emission at 2--5 ACS pixels ($\sim0.05''-0.13''$) from the
center. In particular, the PSF-subtracted ACS images distinctly
appear to have an extended, ring-like echo. Due to limitations of
the image resolution, we cannot confirm any structure or flux within
2 ACS pixels from the SN. The late-time spectrum of SN 2006X can be
reasonably fit with two components: a nebular spectrum of a normal
SN~Ia and a synthetic LE spectrum. Both image and spectral analysis
show a rather blue color for the emission of the LE, suggestive of a
small average grain size for the scattering dust. Using the Cepheid
distance to M100 of 15.2 Mpc, we find that the dust illuminated by
the resolved LE is $\sim$27--170~pc from the SN. The echo inferred
from the nebular spectrum appears to be more luminous than that
resolved in the images (at the $\sim$2$\sigma$ level), perhaps
suggesting the presence of an inner echo at $<$2 ACS pixels
($\sim0.05''$). It is not clear, however, whether this possible
local echo was produced by a distinct dust component (i.e., the
local circumstellar dust) or by a continuous, larger distribution of
dust as with the outer component. Nevertheless, our detection of a
significant echo in SN 2006X confirms that this supernova was
produced in a dusty environment having small dust particles.

\end{abstract}


\section{Introduction}

Light echoes (LEs) are produced when light emitted by the explosive
outburst of some objects is scattered toward the observer by the
foreground or surrounding dust, with delayed arrival time due to the
longer light path. This phenomenon is rare, having been observed
only around a few variable stars in the Galaxy, and around several
extragalactic supernovae (SNe). The best-studied events are SN 1987A
(Schaefer 1987; Gouiffes et al. 1988; Chevalier \& Emmering 1988;
Crotts 1988; Crotts, Kunkel, \& McCarthy 1989; Bond et al. 1990; Xu
et al. 1995) and the peculiar star V838 Mon (Bond et al. 2003).
Other SNe with LEs include the Type II SNe 1993J (Liu et al. 2002;
Sugerman 2003), 2002hh (Meikle et~al. 2006; Welch et al. 2007), and
2003gd (Sugerman 2005; Van Dyk et al. 2006), as well as the Type Ia
SNe 1991T (Schmidt et al. 1994; Sparks et al. 1999), 1998bu
(Cappellaro et al. 2001; Garnavich et al. 2001), and possibly 1995E
(Quinn et al. 2006). Besides their spectacular appearance, LEs offer
a unique means to diagnose the composition, distribution, and
particle size of the scattering dust. In particular, LEs from the
circumstellar environments might provide constraints on SN
progenitors.

The Type Ia SN 2006X was discovered on 2006 February 7.10 (UT dates
are used throughout this paper) by S. Suzuki and M. Migliardi (IAUC
8667, CBET 393) in the nearby spiral galaxy NGC 4321 (M100).
Extensive photometric and spectroscopic coverage is presented by
Wang et al. (2007, hereafter W07). They suggest that SN 2006X is
highly reddened [$E(B - V)_{\rm host} = 1.42 \pm 0.04$ mag] by
abnormal dust with $\Re_{V} = 1.48 \pm 0.06$. Its early-epoch
spectra are characterized by strong, high-velocity features of both
intermediate-mass and iron-group elements. In addition to the
anomalous extinction and the very rapid expansion, SN 2006X exhibits
a continuum bluer than that of normal SNe~Ia. Moreover, its
late-time decline rate in the $B$ band is slow, $\beta = 0.92 \pm
0.05$ mag (100~{\rm d})$^{-1}$, significantly below the 1.4 mag
(100~{\rm d})$^{-1}$ rate observed in normal SNe~Ia and comparable
to the decay rate of 1.0 mag (100~{\rm d})$^{-1}$ expected from
$^{56}$Co $\rightarrow$ $^{56}$Fe decay. This may suggest additional
energy sources besides radioactive decay, such as the interaction of
the supernova ejecta with circumstellar material (CSM) and/or a LE.

Attempts to detect the CSM in SNe~Ia in different wavebands were
unsuccessful before SN 2006X, and only some upper limits could be
placed (see Patat et al. 2007a, and references therein) except for
the peculiar SNe~Ia/IIn 2002ic (Hamuy et al. 2003; Deng et al. 2004;
Wang et al. 2004; Wood-Vasey et~al. 2004) and 2005gj (Aldering
et~al. 2006; Prieto et~al. 2007). Recent progress in this respect
was made from high-resolution spectroscopy by Patat et~al. (2007b,
hereafter P07), who find time-variable Na~I~D absorption lines in
spectra of SN 2006X. This has been interpreted as the detection of
CSM within a few 10$^{16}$ cm ($\sim 0.01$ pc) from the explosion
site of the supernova. With the inferred velocity, density, and
location of the CSM, P07 proposed that the companion star of the
progenitor of SN 2006X is most likely to be a red giant (but see
Hachisu et~al. 2007, who present a main-sequence star model with
mass stripping). Note, however, that SN 2006X exhibited somewhat
abnormal features in spectra and photometry; it may not represent a
typical SN~Ia. Multi-epoch, high-resolution spectral observations of
SN 2007af, a normal SN~Ia, do not reveal any significant signature
of CSM absorption (Simon et~al. 2007).

In this paper we report the discovery of an optical LE around SN
2006X, with evidence from {\it Hubble Space Telescope (HST)}
Advanced Camera for Surveys (ACS) images and Keck optical spectra.
The paper is organized as follows. In \S 2 we briefly describe the
late-epoch data available for SN 2006X, while the data analysis and
the interpretation are presented in \S 3. We discuss the properties
of the light echo and the underlying dust in \S 4. Our conclusions
are given in \S 5.

\section{Observations}

\subsection{{\it HST} Archival Images}
Several epochs of {\it HST} data covering the site of SN 2006X are
publicly available in the MAST archive. The pre-discovery images
were taken on 1993 December 31 (Proposal ID 5195: PI, Sparks) by the
Wide Field Planetary Camera 2 (WFPC2) in the F439W and F555W
filters, with the same integration time of 1800~s. The images taken
prior to the SN explosion allow us to examine the immediate
environment of the progenitor, whereas the post-explosion
observations enable us to search for a possible LE. The most recent,
post-discovery images of SN 2006X were obtained with the High
Resolution Channel (HRC, with a mean spatial resolution of $0.026''$
pixel$^{-1}$) of {\it HST}/ACS on 2006 May 21 (90~d after $B$
maximum) and on 2006 December 25 (308~d after $B$ maximum),
respectively (GO--10991; PI, Arlin Crotts). At $t = 90$~d, the SN
was imaged in F435W (1480~s), F555W (1080~s), and F775W (1080~s),
while at $t = 308$~d, the SN was again observed in the same three
bandpasses, with the exposure times of 920~s, 520~s, and 520~s,
respectively.

The standard {\it HST} pipeline was employed to pre-process the
images and remove cosmic-ray hits. In Figure 1 we show the pre- and
post-explosion {\it HST} images of SN 2006X in the F555W filter.
This pre-discovery image does not reveal any significant source
brighter than 24.0 mag in F555W, excluding the possibility of a
significant star cluster at the location of SN 2006X. Neither of the
two post-discovery images exhibits any resolved LE arcs or rings
around SN 2006X. The magnitudes of SN 2006X were measured from the
{\it HST} ACS images with both the Dolphot method (Dolphin 2000) and
the Sirianni procedure (Sirianni et~al. 2005), and the mean
photometry is given in Table 1.



\subsection{Keck Optical Spectrum}
Observations of nebular-phase spectra provide an alternative way to
explore the possibility of an LE around SNe, as the scattered,
early-phase light will leave a noticeable imprint on the nebular
spectra (e.g., Schmidt et~al. 1994; Cappellaro et~al. 2001) when the
SN becomes dimmer. Two very late-time spectra of SN 2006X taken at t
$\approx$ 277 and 307 days after $B$ maximum were published by W07
(see their Fig. 19), which were obtained by the Keck telescopes at
the W. M. Keck Observatory: one with the Low Resolution Imaging
Spectrometer (LRIS; Oke et al. 1995) mounted on the 10~m Keck I
telescope, and the other with the Deep Extragalactic Imaging
Multi-Object Spectrograph (DEIMOS) mounted on the 10~m Keck II
telescope. In the following analysis we focus on the LRIS spectrum
taken at $t \approx 277$~d because of its wider wavelength coverage.

\section{Data Analysis}
\subsection{Late-Time Light Curves}
Figure 2 shows the absolute $B$, $V$, and $I$ light curves of SN
2006X and SN~Ia 1996X (Salvo et~al. 2001). The former were obtained
using the Cepheid distance $\mu = 30.91 \pm 0.14$ mag (Freedman
et~al. 2001), and corrected for extinction in the Milky Way
($A_{V}({\rm MW})$ = 0.08 mag; Schlegel et~al. 1998) and in the host
galaxy ($A_{V}({\rm host})$ = 2.10 mag; W07). The distance modulus
and host-galaxy extinction for SN 1996X are derived by Wang et~al.
(2006) and Jha et al. (2007) using independent methods, and we
adopted the mean values $\mu = 32.11 \pm 0.15$ mag ($H_{0} = 72$ km
s$^{-1}$ Mpc$^{-1}$ and $A_{V}({\rm host}) = 0.08$ mag are assumed
throughout this paper). The absolute magnitudes of these two SNe are
similar near maximum, except in the $I$ band where SN 2006X is
$\sim$0.4 mag fainter than SN 1996X.

Noticeable differences between the two SNe emerge in $B$ one month
after maximum light, when SN 2006X begins to decline slowly at a
rate of $0.92 \pm 0.05$ mag (100~d)$^{-1}$. The discrepancy reaches
about 0.9 mag in $B$ at $t = 308$~d, while it is $\sim$0.7 mag in
$V$ and $\sim$0.2 mag in $I$. This suggests that the emission of SN
2006X is 130\% $\pm$ 60\% higher in $B$, 90\% $\pm$ 40\% higher in
$V$, and 20\% $\pm$ 20\% higher in $I$ with respect to SN 1996X. The
large error bars primarily reflect the uncertainty in the distances.

The apparently overluminous behavior seen in SN 2006X in the tail
phase is possibly linked to the light scattering of the surrounding
dust, though the interaction of the SN ejecta with the CSM produced
by the progenitor system and/or the excess trapping of photos and
positrons (created in $^{56}$Co $\longrightarrow$ $^{56}$Fe decays
within the ejecta) cannot be ruled out. The resultant LE, if
present, may not be directly resolved even in the {\it HST}/ACS
images at a distance of $\sim$15 Mpc due to the limited angular
resolution. To examine this conjecture, in \S 3.2 we compare the
PSFs of the SN and local stars, and in \S 3.3 we apply the
image-subtraction technique to analyze the SN images.


\subsection{Radial Brightness Profile}

The radial brightness profiles of the images of the SN and local
stars in the same field (see Fig. 1) are compared in Figure 3. These
were obtained by extracting the flux using different apertures,
ranging from 0.1 to 10 pixels with a resolution of 0.1 pixel. The
fluxes of the four local stars labeled in Figure 1 are scaled so
that the integrated flux within the 10-pixel aperture is the same.
Based on the distribution of the radial profiles of these four
stars, we derived a mean radial profile with the same integrated
flux through Monte Carlo simulations. The radial profile of the four
stars was thus normalized by the peak flux of the simulated radial
profile.

One can see that the star profiles are uniform at large radius but
show noticeable scatter within 2 pixels from the center. For
comparison, the central flux of SN 2006X is scaled to be 1.0, with
the assumption that the central region of the SN image was not
affected by any LE. At $t = 90$~d the SN profile does not show a
significant difference from that of the local stars. At $t = 308$~d
the SN profiles appear distinctly broader at radii of 2--4 pixels,
especially in the F435W and F555W images. Note that the SN data are
quite steep at around 1 pixel, probably due to noise.

The inset plot of Figure 3 shows the residual of the radial profile
between SN 2006X and the local stars. This was obtained by
subtracting the simulated radial profile of the stars from that of
the SN. Also plotted is the scatter of the simulated profile of the
local stars. At $t = 308$~d, the SN shows significant extra flux in
the F435W, F555W, and F775W filters at radii of $\sim$2 to 5 pixels,
suggesting the presence of a LE. Such a residual flux was not
present at $t = 90$~d.

One can see some structure (peaks and valleys) at $<$2 pixels in the
inset residual plot of Figure 3. These alternating negative/positive
residuals clearly show that the substructure within the inner 2
pixels cannot be trusted, and could result from the misalignment of
the peak surface brightness of the images. Of course, it is possible
that part of the LE is so close to the SN, but the above analysis
cannot definitively reveal it.

Integrating the overall residual emission in the range 2--10 pixels,
we find that the observed LE brightness is $\sim$22.8 mag in F435W,
$\sim$22.0 mag in F555W, and $\sim$22.1 mag in F775W. Its
contribution to the total flux of the SN + LE is $\sim$29\% in
F435W, $\sim$27\% in F555W, and $\sim$11\% in F775W. In view of
potential additional LE emission at radii $<$2 pixels, these values
are probably lower limits to the true brightness of the LE.


Although Star 1 and SN 2006X show some diffraction spikes in Figure
1, the spikes have the same shape and orientation for all stars in
the field. Thus, (a) they should affect the radial surface
brightness profiles of all stars in the same way, and not affect the
excess light from an echo, and (b) they should be adequately removed
by the image subtraction procedure (\S 3.3).

The difference between the radial profile of SN 2006X and other
stars can also be demonstrated by their measured full width at
half-maximum intensity (FWHM). Table 2 lists the FWHM of SN 2006X
and the average value of several local stars, obtained by running
the IRAF\footnote{IRAF, the Image Reduction and Analysis Facility,
is distributed by the National Optical Astronomy Observatory, which
is operated by the Association of Universities for Research in
Astronomy, Inc. (AURA) under cooperative agreement with the National
Science Foundation (NSF).} ``imexamine'' task in three modes: $r$
(radial profile Gaussian fit), $j$ (line 1D Gaussian fit), and $k$
(column 1D Gaussian fit). At $t = 90$~d, the PSF of SN 2006X is
comparable to that of the average values of the local stars, while
at $t = 308$~d, the SN exhibits a significantly broader profile. The
FWHM increases by about 0.3 pixel in the $r$-profile and by
$\sim$1.0 pixel in the $j$-profile and $k$-profile with respect to
the local stars. The reasonable interpretation is that the PSF is
broadened by scattered radiation (that is, the LE).

\subsection{Light Echo Images}

The radial-profile study suggests the presence of a LE in SN 2006X.
In this section, we apply image subtraction to provide further
evidence for the LE, and study its two-dimensional (2D) structure.

We extract a small section ($20 \times 20$ pixels) centered on SN
2006X and Star 1 (the brightest star in the field), and align their
peak pixels to high precision (0.01 pixel). We then scale Star 1 so
that its peak has the same counts as that of SN 2006X, and subtract
it from the SN 2006X image. The underlying assumption is the same as
in our radial-profile study: the central peak of SN 2006X is not
affected by any LE.

Figure 4 shows the PSF-subtracted images of SN 2006X. The left panel
shows the subtracted images at the original {\it HST}/ACS
resolution. To bring out more details, the middle panel shows
subsampled images by using a cubic spline function to interpolate
one pixel into $8 \times 8$ pixels. The right panel has three
circles (with radii of 2, 4, and 6 pixels, respectively)
overplotted. The residual images all show an extended, bright,
ring-like feature around the supernova, consistent with the general
expected appearance of a LE. These features emerge primarily at
radii of 2--4 pixels (or $0.05''$--$0.11''$) in the images,
consistent with those derived above from the radial profiles. The
central structure seen within a circle of radius 2 pixels (e.g., the
asymmetric feature in F435W, the double features in F775W, and the
arc in F555W) are not to be trusted; due to the limited spatial
resolution, the images used for the image subtraction may not be
perfectly aligned (in terms of the geometry and/or the flux of the
central regions), and some artifacts could be introduced at the
center of the subtracted images. Similarly, the apparent clumps
within the echo ring are not reliable, generally being only a few
pixels in size.

The integrated flux, measured from the PSF-subtracted images at
2--10 pixels from the SN site, contributes to the total flux of SN +
LE by $\sim$33\% in F435W, $\sim$29\% in F555W, and $\sim$9\% in
F775W. This is fully consistent with the above estimate from the
radial-profile analysis, taking into account the uncertainty in the
PSF subtraction. The brightness of the LE component is estimated to
be $\sim$22.7 mag in F435W, $\sim$21.9 mag in F555W, and $\sim$22.3
mag in F775W.

As with the radial-profile analysis, the PSF-subtraction method
might remove some fraction of flux from the LE itself; it had been
assumed that none of the flux in the central 2-pixel radius is
produced by the LE, but this might be incorrect. Thus, our estimate
of the echo flux from the image analysis may be only a lower limit
of the true LE emission. In view of the image analysis, we cannot
verify or rule out that the LE may be distributed continuously from
the SN site to an angular radius of $\sim$6 pixels ($0.15''$).

\subsection{Light Echo Spectrum}

A consistency check for the existence of a LE around a source can
also be obtained by comparing the observed supernova spectrum and
the synthetic spectrum using an echo model. The observed spectrum
should be a combination of the intrinsic late-time SN spectrum and
the early-time scattered SN spectrum. Inspection of the late-epoch
Keck spectrum (see Fig. 19 of W07) clearly reveals that SN 2006X
behaves unlike a normal SN~Ia, showing a rather blue continuum at
short wavelengths and a broad absorption feature near 6100~\AA\
(probably due to Si~II $\lambda$6355).

To construct the composite spectrum containing the echo component,
we use the nebular-phase spectrum of SN 1996X to approximate that of
SN 2006X. SN 1996X is a normal SN~Ia in the elliptical galaxy NGC
5061 (Salvo et~al. 2001), with $\Delta m_{15} = 1.30 \pm 0.05$ mag,
similar to that of SN 2006X (W07). Late-time optical spectra with
wide wavelength coverage and high signal-to-noise ratio (S/N) are
available on day 298 for SN 1996X (Salvo et~al. 2001;
http://bruford.nhn.ou.edu/$^{\thicksim}$suspect/) and on day 277 for
SN 2006X (W07). Comparing the spectrum of SN 2006X obtained at $t =
277$~d with that taken at $t = 307$~d, we found that the overall
spectral slope changed little during this period. We thus could
extrapolate the original nebular spectra $t = 308$~d, a phase when
both SNe have relatively good multicolor photometry. To completely
match the spectrum of SN 2006X, the spectral flux of SN 1996X was
multiplied by a factor of 3.0 caused by the difference in distances.
Extinction corrections have also been applied to the nebular spectra
of these two SNe (W07; Wang et~al. 2006).

We considered the cases of both SN 2006X and SN 1996X as the central
pulse source when deriving the echo spectrum. The observed spectra
of SN 2006X are available at eleven different epochs from about
$-$1~d to 75~d after $B$ maximum, while 14 spectra of SN 1996X are
available from about $-$4~d to 87~d after $B$ maximum (Salvo et~al.
2001). The above spectra were properly dereddened\footnote{Here we
assume that the dust surrounding SN 2006X is a plane-parallel slab
and/or shell, so that both the SN and the LE were affected by
roughly the same amount of extinction.} and interpolated to achieve
uniform phase coverage. Regardless of the original flux calibration,
all of the input spectra have been recalibrated according to their
light curves at comparable phases (W07; Salvo et~al. 2001) and
corrected for the effects of scattering using a similar function,
$S(\lambda) \propto \lambda^{-\alpha}$ (e.g., Suntzeff et~al. 1988;
Cappellaro et~al. 2001). These corrected spectra were then coadded
and scaled, together with the nebular spectrum of SN 1996X, to match
the nebular spectrum of SN 2006X.

The best-fit $\alpha$ values obtained for the combinations of SN
2006X (near $B$ maximum) + SN 1996X (nebular) and SN 1996X (near $B$
maximum) + SN 1996X (nebular) are $3.0 \pm 0.3$ and $3.3 \pm 0.5$,
respectively. One can see that the combination of SN 2006X + SN
1996X gives a somewhat better fit to the observed spectrum of SN
2006X. This is not surprising; the spectrum of SN 2006X differs from
that of a normal SN~Ia at early times, showing extremely broad and
blueshifted absorption minima (W07). The large value of $\alpha$ may
indicate a small grain size for the scattering dust. The composite
nebular spectrum and the underlying echo spectrum are compared with
the observed spectrum of SN 2006X in Figure 5 (upper and middle
panels). Given the simple assumption of the scattering function,
incomplete spectral coverage, and intrinsic spectral difference
between SN 1996X and SN 2006X, the agreement between the observation
and the model is satisfactory, with major features in the spectrum
well matched. This provides independent, strong evidence for the LE
scenario. However, the broad emission peak seen at $\lambda \approx
4300$--4500~\AA\ cannot be reasonably fit by the echo model (see
Fig. 5); this mismatch is probably produced by intrinsic features in
the nebular spectrum of SN 2006X.

\subsection{Light-Echo Luminosity and Color}
We can constrain the properties of the LE and the underlying dust
through the luminosity and colors of the LE. The LE luminosity of SN
2006X has been estimated by analyzing the {\it HST} SN images; it
can also be obtained by integrating the echo spectrum shown in
Figure 5. The magnitudes of the echo given by different methods are
listed in Table 3. For the image-based measurement, the error
accounts only for the scatter of the stellar PSF. On the other hand,
for the spectrum-based measurement, the error primarily consists of
the uncertainties in extinction correction (i.e., $\sim$0.2 mag for
SN 2006X and $\sim$0.1 mag for SN 1996X in the $B$ band) and
distance modulus (i.e., $\sim$0.14 mag for SN 2006X and $\sim$0.15
mag for SN 1996X).

We note that the echo inferred from the spectral fitting seems
somewhat brighter than that revealed by the image analysis: $\delta
m_{F435W} = -0.6 \pm 0.3$ mag. This difference is also demonstrated
in the bottom panel of Figure 5, where the flux ratio of the
inferred echo spectrum and the observed spectrum of SN 2006X is
plotted as a function of wavelength. Overplotted are the ratios
yielded for the photometry of the echo image (circles) and the
spectrophotometry of the echo spectrum (squares) in F435W and F555W,
respectively. Such a discrepancy, at a confidence level of only
$\sim2\sigma$, may suggest that there is some echo emission within a
radius of 2 pixels (21\% $\pm$ 12\% of the total flux of SN + LE in
F435W and 17\% $\pm$ 11\% in F555W) that was not resolved by the
image analysis. Despite this possibility, we must point out that the
echo luminosity derived from the echo spectrum may have an error
that is actually larger than our estimate, since we did not consider
possible uncertainties associated with the spectrum itself and the
simple scattering model adopted in our analysis (see \S 3.4).

Assuming that all of the observed differences between the light
curves and spectra of SN 2006X and SN 1996X at $t = 308$~d are
entirely due to the LE around SN 2006X, we can place an upper limit
on the LE brightness as $21.9 \pm 0.3$ mag in F435W and $21.3 \pm
0.3$ mag in F555W. The magnitudes and the resulting color are not
inconsistent with those presented in Table 3, especially in the case
of the spectral fit which likely takes into account most of the echo
emission. This leaves little room for other possible mechanisms for
the extra emission, suggesting that the echo is the primary cause of
the abnormal overluminosity of SN 2006X at $t = 308$~d.

We find, from analysis of both the {\it HST} images and the nebular
spectrum (see Table 3), that the LE has an average color
(F435W--F555W)$_{\rm echo}$ = $0.8 \pm 0.3$ mag (this roughly equals
$(B -V)_{\rm echo}$ = $0.8 \pm 0.3$ mag), which is much bluer than
the SN color at maximum brightness. The LE is clearly brighter in
bluer passbands than at redder wavelengths (see Fig. 5). Comparing
the colors of the echo and the underlying SN light helps us
interpret the dust, as the color shift depends on the scattering
coefficient and hence on the dimensions of the dust grains (Sugerman
2003a).

Integrating over the entire SN light curve (W07) from about $-$11~d
to 116~d after $B$ maximum yields $(B - V)_{\rm SN} = 1.70$ mag for
the overall emission of SN 2006X. The observed change in color,
$\Delta(B - V) = -0.9 \pm 0.3$ mag, is much larger than the color
shift derived for Galactic dust but is comparable to the change
derived for Rayleighan dust\footnote{The Rayleighan dust consists of
only small particles with grain size $<0.01~\mu$m, and hence has a
scattering efficiency proportional to $\lambda^{-4}$ (Sugerman
2003a).}, $\Delta(B - V)_{\rm max} = -0.96$ mag (Sugerman et~al.
2003b). This is consistent with constraints from the direct spectral
fit, which suggests that the dust has a scattering efficiency
proportional to $\lambda^{-3.0}$. We thus propose that the dust
surrounding SN 2006X is different from that of the Galaxy and may
have small-size grains, perhaps with diameter $\lesssim 0.01~\mu$m,
reflecting the shorter wavelengths of light more effectively.
Smaller dust particles are also consistent with the low value of
$\Re_{V} \approx 1.5$ derived by W07.

\subsection{Dust Distance}

Of interest is the distribution of the dust producing the echo; for
example, it may be a plane-parallel dust slab or a spherical dust
shell. Couderc (1939) was the first to correctly interpret the LE
ring observed around Nova Persei 1901. Detailed descriptions of LE
geometries can also be found in more recent papers (e.g., Sugerman
2003a; Tylenda et~al. 2004; Patat 2005). In general, the analytical
treatment shows that both a dust slab and a dust shell could produce
an echo that is a circular ring containing the source. Assuming that
the SN light is an instantaneous pulse, then the geometry of an LE
is straightforward: the distance of the illuminated dust material
lying on the paraboloid can be approximated as
\begin{equation}
R \approx \frac{{}D^{2}\theta^{2} \mp (ct)^{2}}{2ct},
\end{equation}
where $D$ is the distance from the SN to the observer, $\theta$ is
the angular radius of the echo, $c$ is the speed of light, and $t$
is the time since the outburst. The equation with a minus sign
corresponds to the single dust slab, while the plus sign represents
the case for a dust shell.

As suggested by the analysis of the radial profile and the
PSF-subtracted image of the SN, there is a confirmed LE ring
$\sim0.08''$ away from the SN, with a possible width of
$\sim0.03''$. For this echo of SN 2006X, $ct$ = 0.27 pc, which leads
to $R$ $\approx$ 27--120 pc, consistent with the scale of the ISM
dust cloud. As the dust cloud in front of SN 2006X seems to be very
extended, we do not give the thickness of the dust along the line of
sight. Considering the possible echo emission within 2.0 pixel
($\sim0.05''$) inferred from the echo luminosity (see discussion in
\S 4.1) and that extending up to 5 pixels ($\sim0.13''$; see Fig.
3), the actual distribution of the dust may be from $<$27~pc to
$\sim$170~pc from the SN.

In principle, one can also estimate the distance of the dust itself
from the SN through a fit to the observed echo luminosity using the
light-echo model (e.g., Cappellaro et~al. 2001), as the actual echo
flux is related to the light emitted by the SN, the physical nature
of the dust, and the dust geometry. However, current analytical
treatments for the LE model must assume some idealized
configuration, which may not apply to the dust surrounding SN 2006X
that is found to probably have smaller dust grains with $\Re_{V}
\approx 1.5$ and a relatively extended distribution. Moreover,
multiple scattering processes rather than a single scattering should
be considered in the echo model due to the large optical depth
measured from the dust: $\tau^{V}_{d} \approx 2.0$ for SN 2006X.
Detailed modelling of the LE emission seen in SN 2006X is beyond the
scope of this paper.

\section{Discussion}
Analysis of both the late-time {\it HST} images and the late-time
Keck optical spectrum favors the presence of a LE in SN 2006X, the
fourth non-historical SN~Ia with a detection of echo emission.
Comparison of the SN 2006X echo with the other three known events,
SNe 1991T, 1995E, and 1998bu, shows that the Type Ia echoes may have
a wide range of dust distances from $\lesssim$ 10 pc to $\sim$ 210
pc. The echo detected in SN 1991T is consistent with being a dust
cloud of radius 50 pc (Sparks et~al. 1999), while the echo
speculated from SN 1995E probably corresponds to a dust sheet at a
distance of $207 \pm 35$ pc (Quinn et~al. 2006). Garnavich et~al.
(2001) proposed from the {\it HST} WFPC2 imaging that SN 1998bu may
have two echoes, caused by dust at $120 \pm 15$ pc and $<10$ pc away
from the SN; the outer echo is consistent with an ISM dust sheet,
while the inner component is likely from the CSM dust. On the other
hand, the resolved echo image of SN 2006X appears quite extended in
the direction perpendicular to the line of sight. This yields a dust
distance spanning from $\sim$ 27 pc to $\sim$ 170 pc away from the
site of the SN, indicating that the dust causing the LE may not be a
thin dust sheet but could be a cloud or shell distribution of the
dust around the progenitor or a more complicated dust system.

The echo from SN 2006X is found to be brighter than that of the
other three Type Ia echo events. Assuming the echo magnitude listed
in Table 3 and the SN peak magnitude derived in W07, one can find
that the echo flux with respect to the extinction-corrected peak
magnitude of SN 2006X is $\sim$9.6 mag in $V$. Quinn et~al. (2006)
proposed that all of the other three Type Ia echoes (SNe 1991T,
1995E, 1998bu) show a striking similarity in their echo brightness
relative to the extinction-corrected peak SN brightness, $\Delta V
\approx 10.7$ mag. According to the analytical expression of the
dust scattering (e.g., Patat 2005), the excess echo brightness from
SN 2006X by $\sim$1 mag perhaps suggests a dust distribution closer
to the SN, given the similar optical depth for SNe 2006X and 1995E.
The SN 2006X echo emission also shows a prominent wavelength
dependence, with more light from the shorter wavelengths, suggestive
of smaller-size dust around SN 2006X. This is also demonstrated by
the difference of the scattering coefficient $\alpha$ required to
fit the observed nebular spectrum, which is $\sim$3.0 for SN 2006X,
$\sim$2.0 for 1991T (Schmidt et~al. 1994), and $\sim$1.0 for SN
1998bu (Cappellaro et~al. 2001).

In fitting the nebular spectrum, the echo brightness is found to be
$\sim$ 60\% brighter than that from the echo image at the
$\sim$2$\sigma$ level, likely suggesting the presence of a local
echo that was not resolved at the regions close to the SN site.
Regarding the location of the echo emission in SN 2006X, one may
naturally tie the distribution of the dust underlying the echo to a
combination of local CSM dust and distant ISM dust, given the quite
extended dust distribution and the small dust grains that were not
typical for the ISM dust. Detection of the CSM dust is of particular
importance for understanding SN~Ia progenitor models. P07 recently
reported the detection of CSM in SN 2006X from variable Na~I~D
lines, and they estimate that the absorbing dust is a few $10^{16}$
cm from the SN. It is hence expected that an echo very close to the
SN ($<$0.01~pc away) should be produced, although the SN UV
radiation field could destroy or change the distribution of the
surrounding dust particles out to a radius of a few $10^{17}$ cm
(Dwek 1983). However, it is not possible for us to detect the
emission of such a close CSM echo at $t = 308$~d, since the maximum
delayed travel time of the light for this echo is $<$0.1~yr and the
SN radiation decreases with time.

As noted by W07, the spectrum of SN 2006X probably showed a UV
excess at $t \approx 30$~d. This may be a signature of the nearby
CSM claimed by P07, but the S/N of the spectrum is quite low below
4000~\AA. In this case, the possible echo emission at $<$27~pc
inferred from the nebular spectrum at $t = 308$~d could result from
a dust shell that is farther out than that claimed by P07. This is
possible if the CSM dust around SN 2006X has multiple shells, such
as the dust ring (or shell) of a planetary nebula (Wang 2005) and
nova-like shells.


The presence of a local echo helps explain the slow decline of the
$B$-band light curve of SN 2006X at early phases. Nevertheless, the
local echo (if present) is not necessarily from the CSM dust, as
forward scattering from the distant dust cloud in front of the SN
could also produce an echo of very small angular size. To further
distinguish between the two possible cases of distant ISM plus local
dust and single ISM dust, future {\it HST} observations of SN 2006X
are necessary. More late-phase {\it HST}/ACS images would help
constrain the evolution of the LE. Using equation (1), we can
predict the evolution of the echo ring with time. If the dust formed
as a result of past mass loss from the central source, the echo will
be more symmetric and the expansion will slow down after the
initially rapid phase; with time, its size will eventually shrink to
zero. On the other hand, if the dust is of interstellar origin, the
echo should expand continuously with slowly decreasing brightness as
more-distant regions are illuminated. Assuming that the inner
component of the echo within 2 pixels is caused by a CSM dust shell
$\sim$1~pc from the SN, then the emission within 2 pixels will
finally decrease to zero at $t \approx 6.5$~yr. In contrast, the
local echo from distant ISM dust should remain nearly constant for a
longer time.

It is worth pointing out that the recent nearby SNe~Ia, SN 2007gi
(CBET 1017, CBET 1021), SN 2007le (CBET 1100, CBET 1101), and
probably SN 2007sr (CBET 1172,1174, ATEL 1343), may exhibit
high-velocity features in their spectra similar to those of SN
2006X. If the SN 2006X-like events preferentially occur in
environments with abundant ISM dust or CSM dust (Wang et~al. 2008,
in prep.), then we might expect to detect late-time echo emission in
the above three SNe~Ia. Thus, it would be interesting to obtain
future high-resolution {\it HST}/ACS images of these SNe.

\section{Conclusions}
The emergence of a LE in SN 2006X has been confirmed with
PSF-subtracted {\it HST} ACS images which show a ring-like, but
rather extended, echo 2--5 pixels ($0.05''$--$0.13''$) from the SN
site at $t = 308$~d past maximum brightness. A Keck nebular spectrum
of the SN taken at a similar phase provides additional evidence for
the LE scenario; it can be decomposed into a nebular spectrum of a
normal SN~Ia and a reflection spectrum consisting of the SN light
emitted at early phases.

From the resolved echo image, we derive that the intervening dust is
$\sim$27--170~pc from the supernova. Based on the quite blue color
of the echo, we suggest that the mean grain size of the scattering
dust is substantially smaller than Galactic dust. Smaller dust
particles are also consistent with the low $\Re_{V}$ value obtained
from the SN photometry. Our detection of a LE in SN 2006X confirms
that this SN~Ia occurred in a dusty environment with atypical dust
properties, as suggested by the photometry (W07).

Analysis of the nebular spectrum might also suggest a local echo at
$<$27 pc (or at $<$2 pixels) that is not resolved in the
PSF-subtracted image. This possible local echo is likely associated
with the CSM dust produced by the progenitors, though detailed
modeling of the echo spectrum and/or further high-resolution imaging
are required to test for the other possibilities, such as very
forward scattering by a distant cloud or CSM-ejecta interaction.


\acknowledgments Some of the data presented herein were obtained at
the W. M. Keck Observatory, which is operated as a scientific
partnership among the California Institute of Technology, the
University of California, and the National Aeronautics and Space
Administration (NASA). The Observatory was made possible by the
generous financial support of the W. M. Keck Foundation. This
research was supported by NASA/{\it HST} grants AR--10952 and
AR--11248 from the Space Telescope Science Institute, which is
operated by the Association of Universities for Research in
Astronomy, Inc., under NASA contract NAS5--26555. We also received
financial assistance from NSF grant AST--0607485, the TABASGO
Foundation, the National Natural Science Foundation of China (grant
10673007), and the Basic Research Funding at Tsinghua University
(JCqn2005036).

\clearpage
\begin{table}
\begin{center}
\caption{Late-Time {\it HST} Photometry of SN 2006X.$^a$} {\small
\begin{tabular}{lllccc}
\tableline\tableline
UT Date& JD$-$2,450,000 & Phase (d) &F435W & F555W & F775W \\
\tableline
05/21/2006&3876.0 & +90.0 &18.71(04) &17.36(02) &16.46(02) \\
12/25/2006&4094.0 & +308.0&21.48(06) &20.56(09) &19.69(02)\\
\tableline
\end{tabular}}
\tablenotetext{a}{Uncertainties in hundredths of a magnitude are
given in parentheses.}
\end{center}
\end{table}

\begin{table}
\begin{center}
\caption{FWHM of SN 2006X and Local Stars in {\it HST} Images.}
{\small
\begin{tabular}{lllllc}
\tableline \tableline
 Object  &      $r$(pixel)    &     $j$(pixel)    &   $k$(pixel)   & bandpass\\
\tableline
& &$t = 90$~d & & \\
\tableline
 SN      &    2.21            & 2.61      &  2.51       & F435W  \\
 star    &    2.17$\pm$0.02   & 2.60$\pm$0.03 & 2.60$\pm$0.05 & F435W \\
 SN      &    2.38            & 2.70       & 2.68       & F555W \\
 star    &    2.39$\pm$0.05   & 2.81$\pm$0.04 &2.78$\pm$0.03& F555W \\
 SN      &    2.79            & 2.84       &2.85        &F775W  \\
 star    &    2.82$\pm$0.05   & 2.89$\pm$0.08& 2.89$\pm$0.03&F775W\\
\tableline
& &$t = 308$~d & & \\
\tableline
SN  &        2.50          &     3.85         & 3.36 &  F435W \\
star&        2.21$\pm$0.03 &     2.81$\pm$0.03& 2.51$\pm$0.02 & F435W \\
SN  &        2.62          &     4.00         & 3.69  & F555W \\
star&        2.33$\pm$0.03 &     2.74$\pm$0.02& 2.64$\pm$0.05&F555W\\
SN  &        2.95          &     3.37         & 3.33  & F775W \\
star&        2.79$\pm$0.01 &     2.94$\pm$0.02 &2.83$\pm$0.02 & F775W \\
\tableline
\end{tabular}}
\end{center}
\end{table}

\begin{table}
\begin{center}
\caption{Light Echo of SN 2006X at t = 308~d} {\small
\begin{tabular}{lclc}
\tableline\tableline
Method & F435W (mag) & F555W (mag) & F775W (mag)\\
\tableline
Residual radial profile (2--10 pixels) &22.8$\pm$0.1  & 22.0$\pm$0.3 & 22.1$\pm$0.7 \\
PSF-subtracted image (2--10 pixels)    &22.7$\pm$0.1  & 21.9$\pm$0.3 & 22.3$\pm$0.9 \\
Synthetic echo spectrum (SN 1996X) &22.2$\pm$0.3  & 21.4$\pm$0.3 & \nodata  \\
Synthetic echo spectrum (SN 2006X) &22.1$\pm$0.3&21.5$\pm$0.3 & \nodata\\
\tableline
\end{tabular}}
\end{center}
\end{table}

\clearpage
\begin{figure}
\figurenum{1} \hspace{-0.2cm} {\plotone{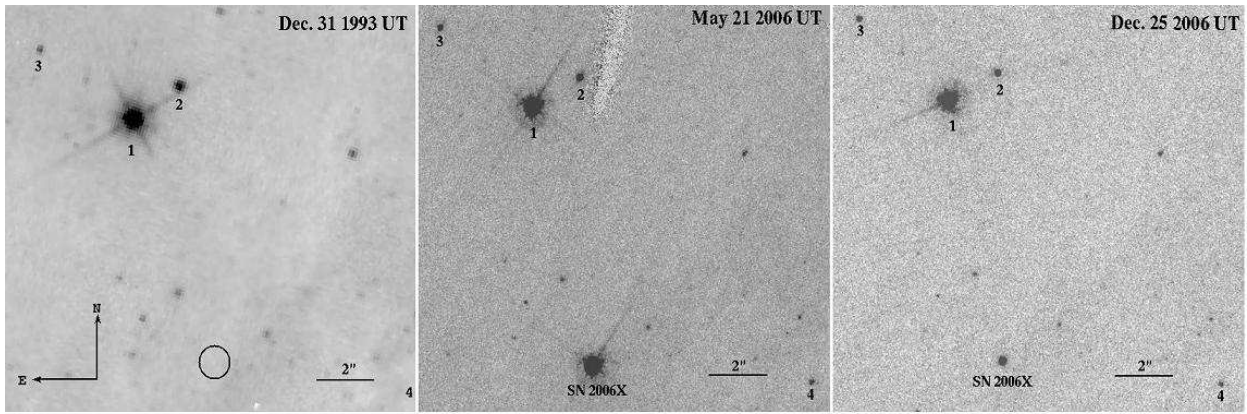}} \hspace{-0.0cm}
\caption{(Left) An {\it HST} image of SN 2006X in the F555W band,
taken on 1993 December 31; the circle marks the position of the SN.
(Middle) The same field was imaged on 2006 May 21 (90~d after $B$
maximum). (Right) The same field was observed on 2006 December 25
(308~d after $B$ maximum). The supernova and the reference stars are
labeled.} \label{fig-1}
\end{figure}

\begin{figure}[htbp]
\figurenum{2} \vspace{-0.5cm} \hspace{-1.0cm} \plotone {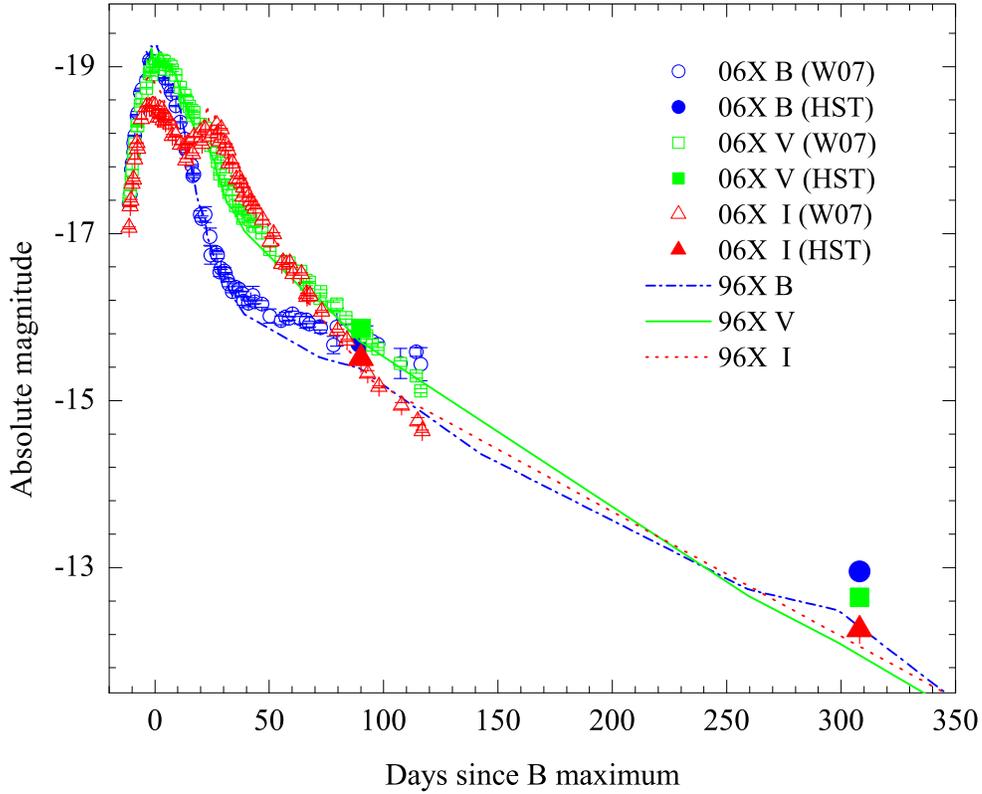}
\caption{Light curve of SN 2006X (symbols) compared with that of
SN~Ia 1996X (dashed lines). The open circles represent the $B$, $V$,
and $I$ data from Wang et~al. (2007). The filled circles denote the
{\it HST} magnitudes that were transformed from the F435W, F555W,
and F775W bands (see Table 1) to the $B$, $V$, and $I$ bands
(respectively) through an empirical correlation (Sirianni et~al.
2005). Proper extinction corrections have been applied to all of the
magnitudes (see text for details).} \label{fig:two} \vspace{-0.5cm}
\end{figure}

\begin{figure}[htbp]
\figurenum{3}\vspace{-0.5cm}\hspace{-1.5cm}
\plotone{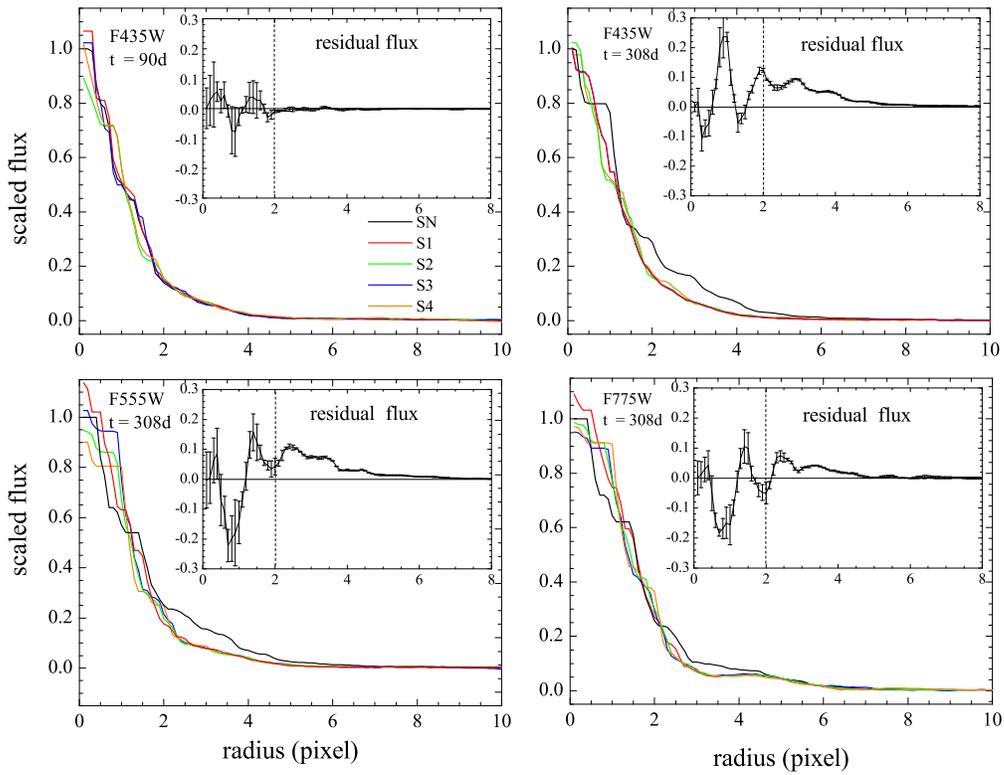} \vspace{-1.0cm} \caption{Radial brightness profile
for SN 2006X in F435W, F555W, and F775W. The black curves show the
SN, while the color curves represent local stars in the same field,
marked in Figure 1 (S1, red; S2, green; S3, blue; S4, orange). The
inset panel shows the residual flux between the SN and the average
of the local stars.} \vspace{-0.0cm} \label{fig:three}
\end{figure}

\begin{figure}[htbp]
\figurenum{4}\vspace{-0.0cm}\hspace{-0.0cm} \plotone{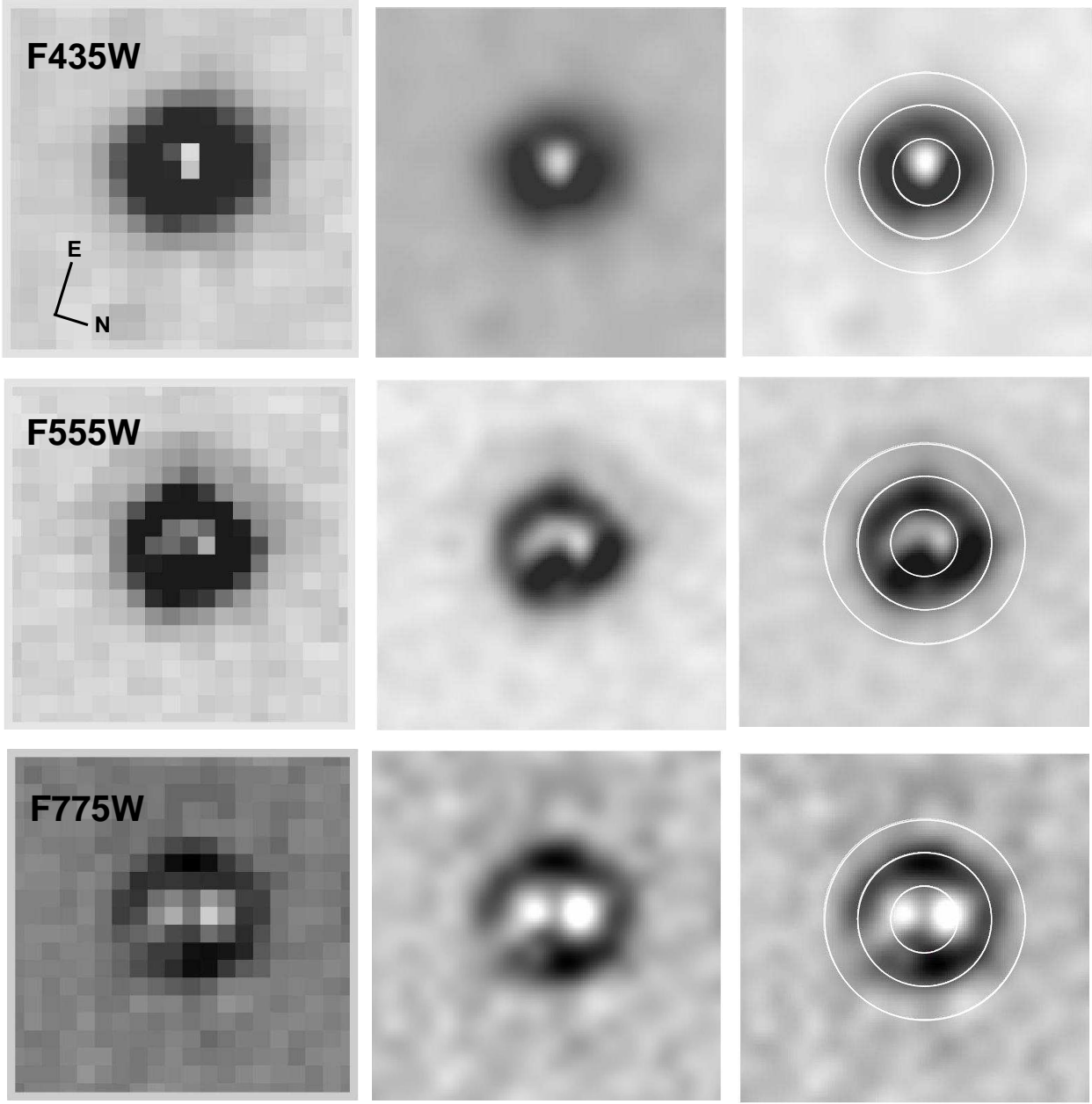}
\vspace{0.0cm} \caption{The PSF-subtracted {\it HST}/ACS images of
SN 2006X (taken on 2006 December 24) with a $0.53'' \times 0.53''$
field surrounding SN 2006X. The supernova is at the center of each
frame. Column (1) shows the residual image of SN 2006X obtained by
subtracting the local bright Star 1 whose central flux is scaled to
that of the supernova; column (2) displays the residual image after
resampling from 1 pixel to $8 \times 8$ pixels; and in column (3)
there are circles of radius 2, 4, and 6 pixels to guide the eye.}
\label{fig:four}
\end{figure}

\begin{figure}[htbp]
\figurenum{5}\vspace{-0.0cm}\hspace{-0.3cm} \plotone{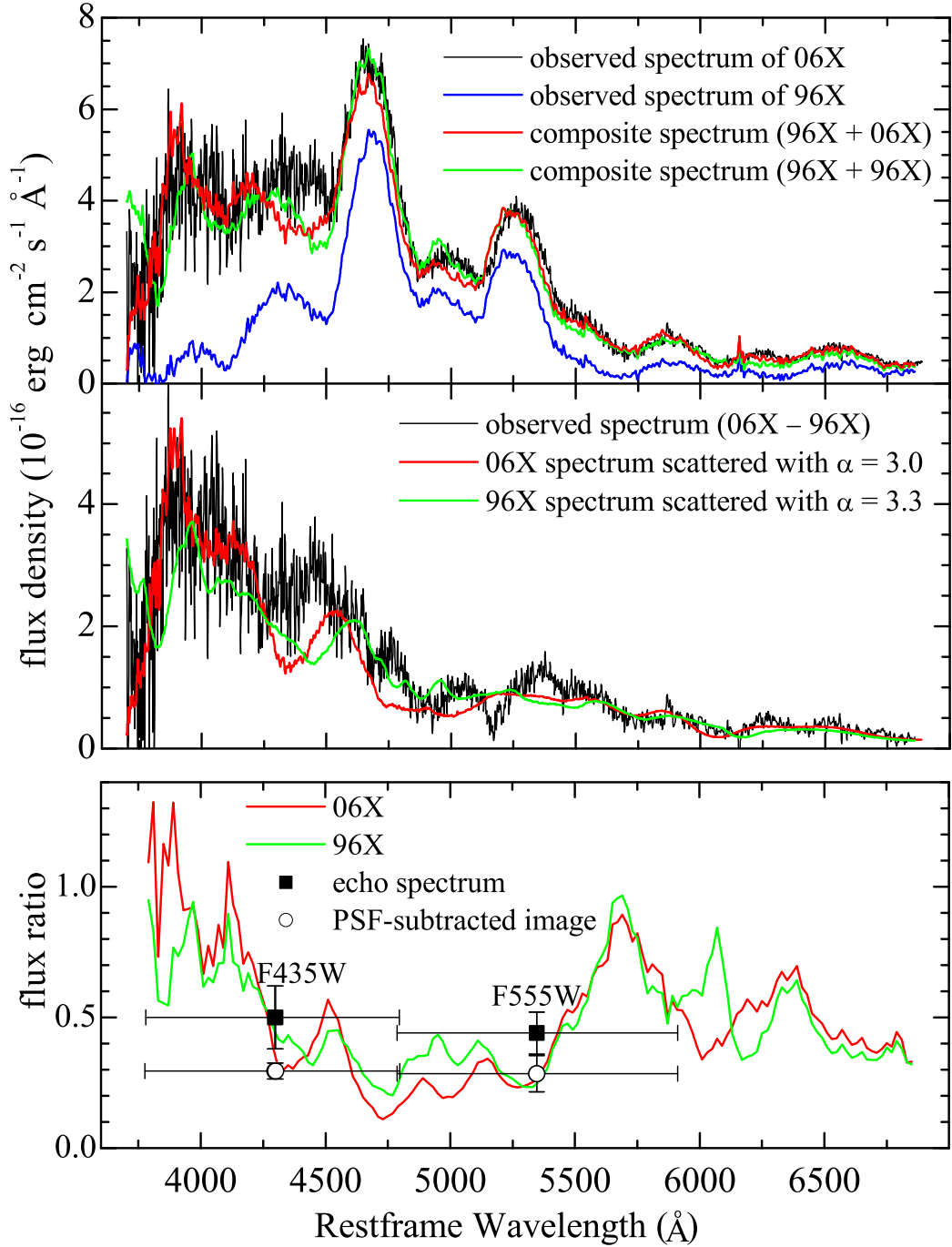}
\vspace{-2.0cm} \caption{Top panel: Comparison of the observed
spectrum of SN 2006X at 308~d with the composite spectrum containing
the nebular emission of SN 1996X and the emission of the echo
computed as described in the text. Middle panel: Residual of the
spectral flux between SN 2006X and SN 1996X at $t = 308$~d,
overlapped with the time-integrated echo spectra. Bottom panel: The
fraction of the total light contributed by the echo as a function of
wavelength. The fractions inferred from the PSF-subtracted image
(open circles) and the spectrophotometry (filled squares) are also
shown with error bars (vertical ones for uncertainties and
horizontal ones for the FWHM of the F435W and F555W filters).}
\label{fig:five}
\end{figure}

\end{document}